# National peer-review research assessment exercises for the hard sciences can be a complete waste of money: the Italian case[1]

*Giovanni Abramo[a,b,\*], Tindaro Cicero[b], Ciriaco Andrea D'Angelo[b]*

[a]Institute for System Analysis and Computer Science (IASI-CNR)
National Research Council of Italy

[b]Laboratory for Studies of Research and Technology Transfer
School of Engineering, Department of Management
University of Rome "Tor Vergata"

**Abstract**

There has been ample demonstration that bibliometrics is superior to peer-review for national research assessment exercises in the hard sciences. In this paper we examine the Italian case, taking the 2001-2003 university performance rankings list based on bibliometrics as benchmark. We compare the accuracy of the first national evaluation exercise, conducted entirely by peer-review, to other rankings lists prepared at zero cost, based on indicators indirectly linked to performance or available on the Internet. The results show that, for the hard sciences, the costs of conducting the Italian evaluation of research institutions could have been completely avoided.

**\* Corresponding author:** Dipartimento di Ingegneria dell'Impresa, Università degli Studi di Roma "Tor Vergata", Via del Politecnico 1, 00133 Rome - ITALY, tel. +39 06 72597362, giovanni.abramo@uniroma2.it

# 1. Introduction

From the 1980s onwards, emphasis on the knowledge economy has encouraged many governments to adopt policies and initiatives to improve the efficiency and effectiveness of their domestic higher education systems. These measures have included the increasing adoption of national research assessment exercises, with key aims of allocating resources by merit and of stimulating increased levels of research productivity on the part of the funding recipients. Historically, the conduct of these evaluation exercises has been founded on peer-review methodology, where research products submitted by institutions are evaluated by appointed panels of experts. In more recent years the development of bibliometric techniques has permitted adoption of metrics that offer support for reviewers in their task of evaluating research products (informed peer-review). In some cases, peer-review for the hard sciences has been completely substituted by bibliometrics, for example in Australia's 2010 ERA exercise.

The pros and cons of peer-review and bibliometric methods have been thoroughly dissected (Horrobin, 1990; Moxham and Anderson, 1992; MacRoberts and MacRoberts, 1996; Moed, 2002; van Raan, 2005; Pendlebury, 2009; Abramo and D'Angelo, 2011). For evaluation of individual scientific products, the literature fails to decisively indicate whether one method is better than the other but demonstrates that there is certainly a positive correlation between peer-review results and citation indicators (Serenko et al, 2011; Aksnes and Taxt, 2004; Oppenheim and Norris, 2003; Rinia et al., 1998; Oppenheim, 1997; Van Raan, 2006), and between peer-review and bibliometric rankings, whether at the individual level (Meho and Sonnenwald, 2000) or for individual organizations (Thomas and Watkins, 1998; Franceschet and Costantini, 2011; Abramo and D'Angelo, 2011).

The severe limits of peer-review emerge when it is applied to comparative evaluation, whether of individuals, research groups or entire institutions. Abramo and D'Angelo (2011b) have contrasted the peer-review and bibliometrics approaches in national research assessments, along the dimensions of accuracy, robustness, validity, functionality, time and costs. Their conclusion is that bibliometric methodology. is by far preferable to informed peer-review. The reason is not the better reliability of citation counts over peer judgment in assessing quality of individual outputs, but rather other implications of the peer-review method. For obvious reasons of costs and time it is clearly unthinkable to utilize peer-review to evaluate the entire output of a national research system. This results in undeniable penalties for peer-review as compared to bibliometric method. First, it prevents any measure of productivity, the quintessential indicator of efficiency for any production system. Second, rankings are sensitive to the size of whatever subset of output is evaluated, which seriously jeopardizes robustness of the peer-review methodology. Third, it implies the necessity of selecting a subset of products (or researchers), which is not necessarily an efficient process, and which can well introduce elements of distortion that compromise the validity of the peer-review method: the resulting rankings then do not reflect the real quality of the subjects evaluated. Fourth, it limits the functionality of the method, since it cannot be applied to all single researchers or research groups. As a consequence, universities do not receive performance rankings of their research staff, to in turn inform their internal selective funding. Finally, with respect to the bibliometric approach, peer-review implies very high costs and long times for execution, which limits its potential frequency of execution.

The effects of the above limits to peer-review methodology have been measured by Abramo et al. (2011), who compared the university performance ranking lists from the first peer-review Italian research assessment exercise (VTR) with those obtained from evaluation simulations conducted with bibliometric indicators. Measurement of the amplitude of shifts revealed notable differences for the rank of the universities involved.

The direct costs of peer-review exercises are very high and vary with the number of products evaluated. For example, the UK's 2008 RAE, which evaluated four outputs per university researcher, cost 12 million pounds[2]. The indirect costs to the institutions evaluated, in terms of opportunity costs for the administrative and research staff time devoted, are estimated at five times the direct costs. In a very current example, the direct costs of the second Italian research assessment exercise are now estimated at 10 to 11 million euros.

The question thus arises as to whether it is worth spending such large amounts of public money to receive rankings that offer such scarce accuracy, when bibliometric techniques would offer more precise and robust results at much lower cost. To offer still more robust evidence of the inadequacy of peer-review for national evaluation of research institutions in the hard sciences, in this work we produce Italian university performance ranking lists based on zero-cost indicators that are indirectly linked to research (such as geographic or economic types), and we draw on rankings lists downloadable for free from the Internet. We compare these ranking lists to the ones from the VTR and from a bibliometric based on the entire scientific production indexed in the Web of Science (WoS), which is our benchmark for accuracy and robustness. The results of the comparison show the paradox that the VTR rankings are not more accurate than the ones available at zero cost.

The first list is a ranking of universities by value of decreasing latitude, from north to south. Two further ranking lists are derived from economic information, specifically GDP per region and regional expenditure on research per inhabitant. Finally we consider two international rankings of universities: one extrapolated from the SCImago Institutions Ranking (SIR) World Report[3] and the other produced by the Italian socio-economic research institute known as CENSIS[4].

In the next section of the paper we illustrate the methodologies used to construct the ranking lists that are the focus of the study. Section 3 shows the analyses carried out, with their results, and Section 4 presents the conclusions.

---

[2]Research Excellence Framework, page 34, downloadable at www.hefce.ac.uk/pubs/hefce/2009/09_38/. Last accessed on Sept. 5, 2012
[3]Available at http://www.scimagoir.com/pdf/sir_2010_world_report.pdf. Last accessed on Sept. 5, 2012
[4] http://www.censis.it/1. Last accessed on Sept. 5, 2012

## 2. Preparation of the various ranking lists

### 2.1 The VTR evaluation

In December 2003 the Italian Ministry for Universities and Research (MIUR) launched its first-ever Research Evaluation exercise, VTR relevant to the period 2001-2003. A national Directory Committee for the Evaluation of Research (CIVR) was made responsible for conducting the VTR. The assessment system was designed to evaluate research and development carried out by 77 universities, 12 public research institutions and 13 private research institutions, the latter participating at their own expense. The present study only deals with the rankings of the universities.

In Italy each university scientist belongs to one and only one specific disciplinary sector (SDS), 370 in all[5], grouped in 14 University Disciplinary Areas (UDAs), 8 of which fall in the hard sciences. CIVR provided the rankings of research performance at the level of single UDAs. As a first step, the CIVR selected panels of experts for each UDA. Universities were then asked to autonomously submit research outputs to the panels: outputs were to be in the proportion of one every four researchers working in the university in the period under observation. Outputs acceptable were limited to articles, books, and book chapters; proceedings of national and international congresses; patents and designs; performances, exhibitions and art works. In the next step, the panels assessed the research outputs and attributed a final judgment for each product, giving ratings of either "excellent", "good", "acceptable" or "limited". The panels were composed of 183 high level peers appointed by the CIVR and called on additional support from outside experts. The judgments were made on the basis of various criteria, such as quality, relevance and originality, international scope, and potential to support competition at an international level. The following quality rating $R$ was used for ranking each university in each UDA[6]:

$$R = \frac{1}{T} \cdot (E + 0.8G + 0.6A + 0.2L) \qquad [1]$$

Where:
$E$; $G$; $A$; $L$= numbers of "excellent, good, acceptable" and "limited" outputs submitted by the university in the UDA
$T$ = total number of outputs submitted by the university in the UDA

A final report ranks universities based on their results under the quality assessment rating. As an example, Table 1 shows the ranking list of the top 10 Italian universities based on $R$, in the UDA "Mathematics and computer science".

The magnitude of the VTR effort is suggested by a few facts: the evaluation included 102 research institutions and examined about 18,000 outputs, drawing on 183 panelists and 6,661 reviewers, with the work taking almost two years and with direct costs mounting to 3.5 million euro.

---

[5] Complete list accessible at http://cercauniversita.cineca.it/php5/settori/index.php, last accessed on Sept. 5, 2012.

[6] http://vtr2006.cineca.it/index_EN.html, last accessed on Sept. 5, 2012.

| University | Selected outputs | E | G | A | L | Rating | Category Ranking (%) |
|---|---|---|---|---|---|---|---|
| Sissa | 3 | 3 | 0 | 0 | 0 | 1.000 | 100 |
| Sannio | 1 | 1 | 0 | 0 | 0 | 1.000 | 100 |
| Rome "Tor Vergata" | 23 | 17 | 5 | 1 | 0 | 0.939 | 96.15 |
| Milan | 28 | 17 | 10 | 1 | 0 | 0.914 | 92.31 |
| Bari Polytechnic | 7 | 4 | 3 | 0 | 0 | 0.914 | 92.31 |
| Milan Polytechnic | 25 | 16 | 7 | 2 | 0 | 0.912 | 90.38 |
| Insubria | 6 | 3 | 3 | 0 | 0 | 0.900 | 86.54 |
| Verona | 4 | 2 | 2 | 0 | 0 | 0.900 | 86.54 |
| Pisa | 42 | 22 | 18 | 2 | 0 | 0.895 | 84.62 |
| Turin Polytechnic | 19 | 9 | 10 | 0 | 0 | 0.894 | 82.69 |

*Table 1: VTR rank list of top 10 Italian universities for UDA mathematics and computer science: E, G, A and L indicate numbers of outputs rated by VTR as excellent, good, acceptable, limited.*

## 2.2 Bibliometric evaluation

The dataset used for the bibliometric analysis is composed of the raw data on the publications (articles, reviews and proceedings papers) listed in the Italian National Country Report extracted from the Thomson Reuters' WoS. Starting from this data, using a complex algorithm for reconciling the authors' affiliation and disambiguating the true author identities, each publication is attributed to the university scientist(s) who produced it (D'Angelo et al., 2011).

The performance assessment was carried out for the eight UDAs[7] where publication in scientific journals is the principle form of codification for research results and where bibliometric measures therefore result as robust. To render greater significance, the field of observation was limited to those SDSs where at least 50% of member scientists produced at least one publication in the period 2001-2003. There are 177 such SDSs, out of the 205 in the hard sciences, representing around two thirds of the overall academic research staff. Over the period examined, these 177 SDSs contained an average of 26,241 scientists distributed in 66 universities (Table 2).

| UDA | SDSs | Universities | Research staff | Publications |
|---|---|---|---|---|
| Agricultural and veterinary sciences | 28 | 36 | 1,910 | 4,175 |
| Biology | 19 | 58 | 4,340 | 14,414 |
| Chemistry | 12 | 57 | 3,065 | 13,017 |
| Earth sciences | 12 | 47 | 856 | 2,131 |
| Industrial and information engineering | 42 | 59 | 3,596 | 13,867 |
| Mathematics and computer sciences | 9 | 56 | 2,197 | 6,384 |
| Medicine | 47 | 51 | 7,925 | 24,152 |
| Physics | 8 | 56 | 2,352 | 12,358 |
| Total | 177 | 66[8] | 26,241 | 78,782* |

*Table 2: Research staff, publications and number of SDSs per UDA; data 2001-2003.*

---

[7] Mathematics and computer sciences; physics; chemistry; earth sciences; biology; medicine; agricultural and veterinary sciences; industrial and information engineering.
[8] To make ranking lists comparable, the dataset has been limited to the 61 universities ranked by SCImago.

Research performance can be calculated at four levels: individual researchers, SDSs, UDAs and universities. The VTR only provides rankings at the university and UDA levels, and for coherence we likewise focus on these two broad groupings. However because of the different intensity of publication across scientific fields, bibliometric comparison of research institution performance should start from field level, i.e. SDS. We therefore take the SDS as base unit of analysis. For measurement of research performance we recur to a labor productivity indicator, named *P*. Measures of productivity are applied to the research staff of every university active in the SDS. Data on staff members of each university and their SDS classifications are extracted from the database on Italian university personnel, maintained by the MIUR[9]. Rather than considering simple output as numerator of the productivity indicator we use the actual outcome, or "impact". As proxy of outcome we adopt the number of citations for the researcher's publications. Abramo et al. (2008) noted though that a number of researchers within the same SDS publish in more than one WoS subject category. This calls for a field (subject category) standardization, when comparing impact of researchers within the same SDS, to account for the varying citation behavior of different subject categories. Citations of a publication are standardized dividing them by the median of citations[10] of all Italian publications of the same year and WoS subject category[11]. Because research collaboration is a growing phenomenon, we extrapolate the real contribution of each researcher to joint production: fractional standardized citations are attributed to a university SDS in function of the co-authors on staff in that SDS relative to the total of co-authors. For the publications in the so-called "life science" categories (corresponding to 66 SDSs of 177), different weights are given to each co-author according to his/her position in the list and the character of the co-authorship (intra-mural or extra-mural)[12]. Finally, values of indicator are averaged on the number of years in which each scientist was officially on staff in Italian universities over the three years under examination, so to have an average annual performance. We assume that production factors other than labor are evenly distributed across universities, which is not far from reality in the Italian public higher education system[13]

At SDS level, in formulae, productivity *P* of university *i* in the SDS *s* is:

$$P_{i,s} = \frac{1}{RS_{i,s}} \cdot \sum_{j=1}^{Ns} \frac{C_j}{C_j^m} \cdot n_{j,i,s} \qquad [2]$$

---

[9]http://cercauniversita.cineca.it/php5/docenti/cerca.php, last accessed on Sept. 5, 2012.
[10]Observed as of 30/06/2009.
[11] For publications in multidisciplinary journals the standardized value is calculated as a weighted average of the standardized values for each subject category.
[12]For the life sciences, position in the list of authors reflects varying contribution to the work. Italian scientists active in these fields have proposed an algorithm for quantification: if the first and last authors belong to the same university, 40% of citations are attributed to each of them; the remaining 20% are divided among all other authors. If the first two and last two authors belong to different universities, 30% of citations are attributed to first and last authors; 15% of citations are attributed to second and last author but one; the remaining 10% are divided among all others. This algorithm could also be adapted to suit other national contexts.
[13] Prior to the VTR, all universities were almost completely financed through non-competitive MIUR allocation.

With:
$n_{j,i,s}$ = fraction of authors of university *i* in SDS *s* on total co-authors of publication *j*, (considering, if publication *j* falls in life science subject categories, the position of each author in the list and the character of the co-authorship, intra-mural or extra-mural).
$N_s$ = total number of publications in SDS *s*
$RS_{i,s}$ = research staff time equivalent of university *i* in SDS *s*, in the observed period
$C_j$ = number of citations received by publication *j*
$C_j^m$ = median of citations received by all Italian publications of the same year and subject category of publication *j*

To calculate the performance at UDA level, we aggregate productivity data of each SDS within the UDA. To account for: i) varying publication and citation intensities of different SDSs and ii) differing representativity, in terms of research staff, of the SDSs present in each UDA, data are conveniently: i) standardized and ii) weighted. At UDA level, in formula, productivity *P* of university *i* in the UDA *u* is:

$$P_{i,u} = \frac{1}{RS_{i,u}} \cdot \sum_{s=1}^{N_u} \frac{P_{i,s}}{\overline{P_s}} \cdot RS_{i,s} \qquad [3]$$

With:
$RS_{i,u}$ = research staff time equivalent of university *i* in UDA *u*, in the observed period
$N_u$ = number of SDSs in the UDA *u*
$\overline{P_s}$ = average value of productivity in SDS *s* of all universities

We apply the same procedure to calculate productivity of the entire university, again beginning from the productivity of each SDS. In formula, productivity *P* per university *i* is:

$$P_i = \frac{1}{RS_i} \cdot \sum_{s=1}^{N_u} \frac{P_{i,s}}{\overline{P_s}} \cdot RS_{i,s} \qquad [4]$$

With:
$RS_i$ = research staff time equivalent of university *i* in the observed period
$N_u$ = number of SDSs where the university is active
$\overline{P_s}$ = average value of productivity in SDS *s* of all universities

## 2.3 Evaluation based on geographic and economic indicators

To generate zero-cost rankings to compare with those from the VTR and bibliometrics, we draw on economic and geographic information concerning the universities and their base regions. Intentionally, the rankings produced here are not based on the quality of research output.

The literature includes studies that show the relative performance of the various European regions in terms of articles cited. For example Bornmann and Leydesdorff (2011) show that, in Italy, the cities with highly cited articles (top 10%) in information science are particularly concentrated in the north of the nation. These results generated the idea of producing a ranking list of Italian universities in decreasing order of the numeric latitude of each home city.

Various studies also show the link between R&D expenditures and development at the regional level. Taking particular note of Guisan (2005), on the positive effects of university research expenditures on regional development, we have ranked universities by university research expenditure per inhabitant in the relative home regions. We produce a third ranking list on the basis of GDP per inhabitant in the university home regions. For both of these lists we draw on data available from the Italian National Institute for Statistics[14] (ISTAT) concerning the year 2002, the middle year of the triennium under study 2001-2003.

**2.4 Evaluations available on Internet for free**

Besides exploring the above indirect indicators of performance, we also examine the possibilities for using the yearly international university "league tables" (e.g. QS World University Rankings, Times Higher Education World University Rankings; The Leiden Rankings; ARWU; SCImago, etc). The most extensive free list is SCImago Institutions Ranking (SIR) World Report, which classifies 2500 universities per year[15] based on publication data from Scopus (Elsevier), with the results on Internet. Since the reports available do not extend as far back as 2003 we refer to the one closest to our period of observation, based on research output from 2004-2008. As justification, we verify that the ranking lists for bibliometric performance P for the period 2001-2003 (see Section 2.2) is strongly correlated (Spearman coefficient: +0.867, p-value= 0.000) to the P-based ranking for 2004-2008. From the SIR web site, we extract the ranks of Italian universities according to the indicators IC, Q1 and NI, where:
- IC, or International Collaboration, represents the percentage of publications co-authored with foreign organizations on total publications of the university.
- Q1, represents the ratio of scientific publications that a university manages to publish in the 25% of the most influential journals according to the SCImago ranking.
- NI, or Normalized Impact, represents the overall scientific impact of institutions. "Normalized Impact values show the ratio between the average scientific impact of an institution and the world average impact of publications of the same time frame, document type and subject area. Normalized Impact is computed using the methodology established by the Karolinska Intitutet in Sweden where it is named "*item oriented field normalized citation score average*"[16].

---

[14]http://www.istat.it/it/ Last accessed on Sept. 5, 2012
[15]SCImago 2010 World Report, available at http://www.scimagoir.com/pdf/sir_2010_world_report.pdf. Last accessed on Sept. 5, 2012.
[16]SCImago metholodgy available at http://www.scimagoir.com/methodology.php?page=indicators#.Last accessed on Sept. 5, 2012.

Finally, we use the reports on individual Italian university faculties published by CENSIS (Center for Social Investments Studies). These offer rankings by various performance indicators, from which we choose the list for the "Research" indicator, referring to competitive funding received by universities from the MIUR's PRIN programme for research projects of national interest. The reference VTR and P rankings are by UDA, therefore we take the CENSIS rankings of the individual faculties and aggregate them into the corresponding UDAs (Table 3).

| Faculty | UDAs |
|---|---|
| Engineering | Industrial and Information engineering |
| Medicine | Medicine |
| Agricultural sciences | Agricultural and Veterinary Sciences |
| Veterinary sciences | Agricultural and Veterinary Sciences |
| Natural and formal sciences | Biology |
| | Chemistry |
| | Physics |
| | Earth Science |
| | Mathematics and computer sciences |

*Table 3: Relation of UDAs to faculties*

## 3. Results and analysis

This section presents the comparisons of the VTR and "zero-cost" rankings to the benchmark P ranking. Table 4 provides a brief review of all the rankings lists with their acronyms and sources and levels of data agglomeration.

| Acronym | Ranking by | Source of data | Level of analysis |
|---|---|---|---|
| VTR | VTR peer-review rating | CIVR | UDA/University |
| P | Biliometric rating | Web of Science | UDA/University |
| LAT | Latitude of university | Google maps | University |
| EXP | Research expenditure by universities (of a region) per inhabitant | ISTAT | University |
| GDP | Gross Domestic Product (of a region) per capita | ISTAT | University |
| IC | Rate of International Collaboration (%) | SCImago | University |
| Q1 | Publications ratio in top journals (%) | SCImago | University |
| NI | Normalized impact | SCImago | University |
| RES | Funding received from PRIN program | CENSIS | Merged UDAs |

*Table 4: Acronyms of rankings used in the analysis*

### 3.1. Rank correlations

To measure the level of accuracy, we calculate the Spearman rank correlations for each ranking. The report is divided in subsections, the first concerning the geographic and economic rankings, second the SCImago rankings, and finally the CENSIS ranking.

*3.1.1 Geographic and economic rankings*

Here we present the results from the correlation analyses relative to university bibliometric rank calculated on the basis of P, the VTR rankings, and rank from geographic and economic information, i.e. latitude (LAT), regional expenditures on R&D (EXP) and gross domestic product per capita (GDP).

We first carry out the correlations without distinguishing by UDA, since not all the rankings provide this level of detail.

From Table 5, we see that there is a significant level of correlation between P rankings and LAT rankings ($\rho$=+0.6291), confirming the strong positive correlation[17] between university performance and latitude of the home city. These values are actually greater than the ones between P and VTR rankings (+0.6162).

The correlations to the two economic variables, EXP and GDP, are weaker, but still significant, at approximately +0.29 and +0.35.

|     | P       | VTR     | LAT     | EXP    | GDP    |
| --- | ------- | ------- | ------- | ------ | ------ |
| P   | 1.0000  |         |         |        |        |
| VTR | 0.6162* | 1.0000  |         |        |        |
| LAT | 0.6291* | 0.5903* | 1.0000  |        |        |
| EXP | 0.2904* | 0.1289  | 0.0584  | 1.0000 |        |
| GDP | 0.3511* | 0.1639  | 0.3549* | 0.1120 | 1.0000 |

*Table 5: Spearman correlation matrix among P, VTR and geographic and economic university rankings*
*\* p-value< 0.05*

Focusing only on the ranking by variable with greatest correlation, i.e. latitude, we carry out further analyses at the UDA level. The results (Table 6) show a certain variability across UDAs and offer two points of reflection. First, the correlation values per UDA are all lower than the overall value per university; second, correlation still remains strong for the biomedical area. The highest value of correlation between LAT and P is seen in Medicine ($\rho$ =+0.6081). Also, for four UDAs out of eight, the value of correlation between P and LAT is significantly higher than that between P rank and VTR rank by peer-review. For example in Agricultural and veterinary sciences the value of correlation for LAT to P is +0.4974, compared to +0.4033 for VTR to P. In earth sciences, where the correlation between ranks by VTR and P is not significant ($\rho$ =+0.1771), the correlation for LAT to P ranks is noticeably higher, and significant ($\rho$ =+0.4642). The correlations for the remaining cases are quite similar except for Industrial and information engineering and Chemistry, where correlation between the P and VTR ranks is decidedly higher ($\rho$ =+0.4920).

---

[17] We followed the guidelines by Cohen (1988) on the strength of association.

|  | *Agricultural and veterinary Sciences* | | | | *Biology* | | |
| --- | --- | --- | --- | --- | --- | --- | --- |
|  | P | VTR | LAT |  | P | VTR | LAT |
| P | 1.0000 |  |  | P | 1.0000 |  |  |
| VTR | 0.4033* | 1.0000 |  | VTR | 0.5267* | 1.0000 |  |
| LAT | 0.4974* | 0.3725* | 1.0000 | LAT | 0.4646* | 0.3908* | 1.0000 |
|  | *Chemistry* | | | | *Earth Sciences* | | |
|  | P | VTR | LAT |  | P | VTR | LAT |
| P | 1.0000 |  |  | P | 1.0000 |  |  |
| VTR | 0.4920* | 1.0000 |  | VTR | 0.1771 | 1.0000 |  |
| LAT | 0.3063* | 0.2601 | 1.0000 | LAT | 0.4642* | 0.1213 | 1.0000 |
|  | *Industrial and information engineering* | | | | *Mathematics and computer sciences* | | |
|  | P | VTR | LAT |  | P | VTR | LAT |
| P | 1.0000 |  |  | P | 1.0000 |  |  |
| VTR | 0.4250* | 1.0000 |  | VTR | 0.4486* | 1.0000 |  |
| LAT | 0.4279* | 0.4822* | 1.0000 | LAT | 0.3241* | 0.3885* | 1.0000 |
|  | *Medicine* | | | | *Physics* | | |
|  | P | VTR | LAT |  | P | VTR | LAT |
| P | 1.0000 |  |  | P | 1.0000 |  |  |
| VTR | 0.4663* | 1.0000 |  | VTR | -0.0389 | 1.0000 |  |
| LAT | 0.6081* | 0.4125* | 1.0000 | LAT | 0.1469 | 0.2578 | 1.0000 |

***Table 6: Spearman correlation matrix among P, VTR and geographical ranking by UDA***
*\* p-value< 0.05.*

### *3.1.2 SCImago ranking*

This second section presents analysis of correlations with ranks derived from the SCImago rankings. In Table 7 we see that the NI ranking is highly correlated with P ranking ($\rho=+0.7225$) and the VTR and P rankings are also correlated but with lower intensity ($\rho=+0.6162$). The correlation between the IC and P ranking is also strong but with still lower intensity ($\rho=+0.5219$).

Given the results showing higher correlation for the NI indicator, we decided to carry out a more detailed analysis of the relationship at the UDA level. Again in this case, since SCImago does not provide rankings by area, we assign the same value of normalized impact to every UDA in each ranked university. The results continue to be significant, with very high correlation values for all UDAs. The highest value is obtained in Agricultural and veterinary sciences ($\rho=+0.6623$), much above the correlation obtained with VTR rank ($\rho=+0.4033$). For the Earth sciences area the correlation with NI is +0.4142, while correlation with VTR rank is not significant. For the Physics area the correlation is not significant for both NI and VTR (Table 8).

|     | P       | VTR     | IC      | Q1      | NI     |
| --- | ------- | ------- | ------- | ------- | ------ |
| P   | 1.0000  |         |         |         |        |
| VTR | 0.6162* | 1.0000  |         |         |        |
| IC  | 0.5219* | 0.5889* | 1.0000  |         |        |
| Q1  | 0.2572* | 0.5026* | 0.3367* | 1.0000  |        |
| NI  | 0.7225* | 0.6691* | 0.5210* | 0.3973* | 1.0000 |

*Table 7: Spearman correlation matrix among P, VTR and SCImago rankings*

| | Agricultural and veterinary Sciences | | | | Biology | | |
| --- | --- | --- | --- | --- | --- | --- | --- |
|     | P       | VTR     | NI     |     | P       | VTR     | NI     |
| P   | 1.0000  |         |        | P   | 1.0000  |         |        |
| VTR | 0.4033* | 1.0000  |        | VTR | 0.5267* | 1.0000  |        |
| NI  | 0.6623* | 0.4557* | 1.0000 | NI  | 0.4981* | 0.5041* | 1.0000 |
| | Chemistry | | | | Earth Sciences | | |
|     | P       | VTR     | NI     |     | P       | VTR     | NI     |
| P   | 1.0000  |         |        | P   | 1.0000  |         |        |
| VTR | 0.4920* | 1.0000  |        | VTR | 0.1771  | 1.0000  |        |
| NI  | 0.3986* | 0.2234  | 1.0000 | NI  | 0.4142* | 0.1005  | 1.0000 |
| | Industrial and information engineering | | | | Mathematics and computer sciences | | |
|     | P       | VTR     | NI     |     | P       | VTR     | NI     |
| P   | 1.0000  |         |        | P   | 1.0000  |         |        |
| VTR | 0.4250* | 1.0000  |        | VTR | 0.4486* | 1.0000  |        |
| NI  | 0.4800* | 0.5104* | 1.0000 | NI  | 0.3935* | 0.5230* | 1.0000 |
| | Medicine | | | | Physics | | |
|     | P       | VTR     | NI     |     | P       | VTR     | NI     |
| P   | 1.0000  |         |        | P   | 1.0000  |         |        |
| VTR | 0.4663* | 1.0000  |        | VTR | -0.0389 | 1.0000  |        |
| NI  | 0.5483* | 0.4921* | 1.0000 | NI  | 0.0784  | 0.1832  | 1.0000 |

*Table 8: Spearman correlation matrix among P, VTR and NI SCImago ranking, by UDA*

### *3.1.3 CENSIS ranking*

This final subsection concerns the analysis of correlation between P rank and the ranks derived from CENSIS. In this case the ranks are presented by macro-UDA, defined in Table 3, since CENSIS ranks are given only by faculty.

The biomedical area, in this case included in the Natural and formal sciences macro-UDAs, continues to show high correlation values (Table 9). For this macro-UDA the correlation between indicator RES and P is actually strong ($\rho=+0.6342$), greater than that between P and VTR ($\rho=+0.5930$). Among the remaining areas, for Engineering and agricultural and veterinary sciences the correlation with the CENSIS ranking is low and not significant.

|       | Agricultural and veterinary sciences | | |       | Natural and formal sciences | | |
|-------|---------|---------|--------|-------|---------|---------|--------|
|       | P       | VTR     | RES    |       | P       | VTR     | RES    |
| P     | 1.0000  |         |        | P     | 1.0000  |         |        |
| VTR   | 0.4060* | 1.0000  |        | VTR   | 0.5930* | 1.0000  |        |
| RES   | 0.1717  | 0.3608  | 1.0000 | RES   | 0.6342* | 0.5043* | 1.0000 |
|       | Medicine | | |       | Engineering | | |
|       | P       | VTR     | RES    |       | P       | VTR     | RES    |
| P     | 1.0000  |         |        | P     | 1.0000  |         |        |
| VTR   | 0.5527* | 1.0000  |        | VTR   | 0.4553* | 1.0000  |        |
| RES   | 0.5819* | 0.4921* | 1.0000 | RES   | 0.2871  | 0.0050  | 1.0000 |

*Table 9: Spearman correlation matrix among P, VTR and CENSIS ranking, by macro-UDA*

### 3.2 Distributions of changes in rank

The results of the Spearman correlation analyses evidence strong association between the P rankings and those derived from informed rankings, particularly from SCImago's NI normalized impact value. However a correlation value around 0.6 does not provide any specific information on the behavior of specific categories of universities, such as those classified at the top of the ranking. Thus we classify the institutions into four classes, as is common in research assessment exercises, according to their percentile ranking. We assign values of 4, 3, 2 and 1, corresponding to the first, second, third and fourth quartiles for the performance value. We then calculate the distribution of the shifts in quartile when a university "shifts" from the P ranking to SCImago NI and peer-review VTR ranking. Table 10 shows that, in comparing the NI ranking to the benchmark bibliometric P, 54.10% of universities remain in the same performance quartile[18]; 32.79% move one quartile and 13.11% make a two quartile jump. No university makes a three quartile jump. However, comparing the VTR and P ranking, the distribution of shifts is slightly different, showing a longer tail to the right: less universities maintain a constant quartile (45.90%) and a greater number shift one quartile (40.98%); 1.64% of the universities actually make the maximum shift of three quartiles.

|         | NI vs P | NI vs VTR | VTR vs P |
|---------|---------|-----------|----------|
| Changes | Relative frequency distributions | | |
| 0       | 54.10%  | 47.54%    | 45.90%   |
| 1       | 32.79%  | 39.34%    | 40.98%   |
| 2       | 13.11%  | 13.11%    | 11.48%   |
| 3       | 0.00%   | 0.00%     | 1.64%    |
|         | Cumulative frequency distributions | | |
| ≤ 0     | 54.10%  | 47.54%    | 45.90%   |
| ≤ 1     | 86.89%  | 86.89%    | 86.89%   |
| ≤ 2     | 100.00% | 100.00%   | 98.36%   |
| ≤ 3     | 100.00% | 100.00%   | 100.00%  |

*Table 10: Distributions of change in rank among NI, VTR and P.*

---

[18] We note that the value change of rank within quartiles for any universities which do not shift quartile, may be larger than that of universities that shift quartile.

This first analysis of the quartile shift distributions indicates that the VTR is less reliable than the NI ranking. Still, we would like more detailed examination of the case of the universities ranked as "top" by P ranking. Table 11 presents various simulations for extraction of top universities for the first two quartiles of the P ranking, with calculation of the numbers of universities that would not register as such under NI or VTR rank. For example, if we consider the top 25% of universities, i.e. those that fall in performance quartile 4, we observe that 7 universities out of 15 (46.7%) are not classified as such for the VTR. This number falls to 4 out of 15 (26.7%) if we consider the ranking for NI.

In general, for all simulations of extracting top universities that we try, from top 5% to top 50%, we observe that with the exception of three cases, the NI rank is always closer than the VTR to the more trustworthy P rank. Only in the case of the extracting the top 5% by P do we obtain superimposition of the rank from VTR (0 variations out of three universities), but given the very slim number of cases that compose this group the data has little relevance.

|  | NI vs P | | VTR vs P | |
| --- | --- | --- | --- | --- |
| Top universities | Variations | Percentage | Variations | Percentage |
| 5% | 2 out of 3 | 66.67% | 0 out of 3 | 0.00% |
| 10% | 2 out of 6 | 33.33% | 3 out of 6 | 50.00% |
| 15% | 2 out of 9 | 22.22% | 5 out of 9 | 55.56% |
| 20% | 4 out of 12 | 33.33% | 5 out of 12 | 41.67% |
| 25% (4$^{th}$ quartile) | 4 out of 15 | 26.67% | 7 out of 15 | 46.67% |
| 30% | 4 out of 18 | 22.22% | 8 out of 18 | 44.44% |
| 35% | 6 out of 21 | 28.57% | 8 out of 21 | 38.10% |
| 40% | 6 out of 24 | 25.00% | 8 out of 24 | 33.33% |
| 45% | 9 out of 27 | 33.33% | 7 out of 27 | 25.93% |
| 50% (3$^{rd}$ quartile) | 9 out of 30 | 30.00% | 7 out of 30 | 23.33% |

*Table 11: Percentile change in rank among NI, VTR and P*

**4. Conclusions**

In previous works the authors had demonstrated the limits in planning of the Italian national research assessment exercise and, for the hard sciences, the undisputable superiority of bibliometrics over the peer-review methodology adopted. In this work we have asked whether it would be possible to prepare university ranking lists at zero cost and with levels of accuracy comparable or superior to those of the VTR ranking list.

Taking a ranking of Italian universities by decreasing latitude from north to south, we found it comparable to the VTR: for the individual disciplines, the results actually showed greater accuracy than the VTR in half the disciplines and lesser accuracy in three out of eight. Again at the level of discipline, the freely-available CENSIS rankings were also equivalent to those from the VTR.

However when we compared SCImago university rankings by average citation impact we found that these lists outperformed the VTR, both at the absolute level and by discipline.

The moral: not only would the application of bibliometric techniques be more precise, more robust and notably less expensive, but the entire direct and indirect costs of the VTR for the hard sciences could be completely avoided by resorting to zero-cost rankings, and these would give results of equal or greater level of accuracy.

Governments in general, and especially the Italian government, should question the competencies of those who are planning national evaluation exercises, or at least ensure that there is a sufficient exchange of knowledge between scholars and practitioners to ensure maximum efficiency, effectiveness and fairness in their conduct.